# Recommendation systems: a joint analysis of technical aspects with marketing implications

Draft version

**Vafopoulos Michalis**

Mathematics Department, Aristotle University of Thessaloniki, Greece

**Oikonomou Michael**

Mathematics Department, Aristotle University of Thessaloniki, Greece

**Abstract** In 2010, Web users ordered, only in Amazon, 73 items per second and massively contribute reviews about their consuming experience. As the Web matures and becomes social and participatory, collaborative filters are the basic complement in searching online information about people, events and products.

In Web 2.0, what connected consumers create is not simply content (e.g. reviews) but context. This new contextual framework of consumption emerges through the aggregation and collaborative filtering of personal preferences about goods in the Web in massive scale. More importantly, facilitates connected consumers to search and navigate the complex Web more effectively and amplifies incentives for quality.

The objective of the present article is to jointly review the basic stylized facts of relevant research in recommendation systems in computer and marketing studies in order to share some common insights.

After providing a comprehensive definition of goods and Users in the Web, we describe a classification of recommendation systems based on two families of criteria: how recommendations are formed and input data availability. The classification is presented under a common minimal matrix notation and is used as a bridge to related issues in the business and marketing literature. We focus our analysis in the fields of one-to-one marketing, network-based marketing Web merchandising and atmospherics and their implications in the processes of personalization and adaptation in the Web. Market basket analysis is investigated in context of recommendation systems. Discussion on further research refers to the business implications and technological challenges of recommendation systems.

# Preface

Searching, social networking, recommendations in various forms, blogging and micro-blogging have become part of everyday life whilst the majority of business applications have migrated to the Web. Understanding and modeling this enormous impact of the Web in macro (e.g. [1]) and micro scale (e.g. [2], [3]) has become a major task for computer and social scientists. The trans-disciplinary field in this direction has been entitled "Web Science" and is focused in the significant reciprocal relationship among the social interactions enabled by the Web's design, the scalable and open applications development mandated to support them, and the architectural and data requirements of these large-scale applications [4], [5], [6].

The Web "curves" physical time and space by adding flexibility, universality [7] and more available options [8], [9] and sources of risks [10]. At the current Web 2.0 era, Users can easily edit, interconnect, aggregate and comment text, images and video in the Web. Most of these opportunities are engineered in a distributed and self-powered level.

In particular, recommendation systems have become mainstream applications in the Web with massive User participation affecting an important part of offline and online industries. During the last twenty years, research and practice on recommendation systems is growing in an increasing pace. This massification creates new business opportunities and challenging research issues in software development, data mining, design of better algorithms, marketing, management and related issues. User and business demands are now setting part of the research agenda in recommendation systems literature. Recently, new research communities (e.g. network analysis) from diverse fields have started to involve in the research of recommendation systems in order to understand the economic behavior of online consumers and its implications to business process and competition.

Computer science literature and related fields are often enriched by bibliographic reviews on the advancements of recommendation systems ([11] is the most recent). To the best of our knowledge, it does not exist an effort to jointly review the technical and business aspects of recommendation systems. Thus, the objective of the present article is to overview the main aspects of relevant research in recommendation systems both in computer and marketing studies in order to create a bridge and facilitate the sharing of common insights.

The article is organized as follows. The first section is devoted in the description of the fundamental changes that the Web brings in the economy. Specifically, the role of recommendation systems is identified as an important part in the transition to more energetic and interdependent consumption patterns. The second section provides an overview of the technical aspects characterizing recommendation systems. After providing a targeted and comprehensive definition of goods and Users in the Web, we describe a classification of recommendation systems based on two families of criteria: how recommendations are formed and input data availability. The classification is presented under a common minimal matrix notation. The third section reviews the related issues of recommendation systems in the business and marketing literature. We focus our analysis in the fields of one-to-one marketing, network-based marketing, Web merchandising and atmospherics and their implications in the processes of personalization and adaptation in the Web. Market basket analysis is investigated in the context of recommendation systems. The final section discusses issues for further research.

# 1 Consumption in the Web

### 1.1 Introduction

Some economists expected that the Web would gradually lead to perfect information in consumption, acute price competition and pricing at the marginal cost followed by low dispersion [12]. The basic arguments were based on lower search and fixed costs, less product differentiation and "frictionless commerce" via the Web. There is not strong evidence that many things have changed in these directions in the markets of ordinary goods, since online prices are still dispersed, not much lower than offline (see for instance [13] and [14]) and many sectors continue to share oligopolistic characteristics.

But what actually changed, and not expected at all, was the emergence of new types of consumption and production, new service sectors (e.g. Software as a Service) and the transformation of existing industries (e.g. mass media). The resulting reconfigurations in the triptych of production-exchange-consumption stemmed from an update in the fundamentals of the economy that the Web brings. Basically, the Web is contributing one major new source of increasing returns in the economy: *more choices with less transaction costs in production and consumption.*

This source of value arises from the orchestration of digital and network characteristics of goods in the Web. More choices in consumption are ranging from larger variety of available goods, to online consumer reviews, recommendations and adaptive content. This updated mode of *connected consumption* allows consumers to make more informed decisions and provides them with stronger incentives to take part in the production and exchange of mainly information-based goods. On the other hand, the provision of more choices with less transaction cost in consumption is not always coming without compensation. The leading native business model in the Web is the forced joint consumption of online information and contextual advertisements in massive scale.

Turning in the *production* side, many business operations virtualized, went online and become less hierarchical, niche online markets and services emerged and traditional industries revolutionized. Decentralized Peer production through loosely affiliated self-powered entities is based on a broader baseline of input and output to create a larger range of possibilities for both producers and consumers [15]. Moreover, the recent emergence of "social commerce" as a consumer-driven online marketplace of personalized, individual-curated shops that are connected in a network, demonstrates the volatile boundaries among production, exchange and consumption in the Web.

**1.2 More energetic and connected consumption**

Due to the rapid penetration of the Web in many technological platforms (e.g. mobile, TV) and social aspects, electronic commerce has become a major activity in ordinary business operations. Almost every firm in the developed world has online presence that describes or/and provides its goods to potential customers. The migration of many business functions in the Web decreased operational costs, primarily, for service-oriented companies. Online commerce is one of the basic components of the Web economy and is gradually becoming an important sector for the entire economy. The Census Bureau of the Department of Commerce announced that the estimate of U.S. retail e-commerce sales for the first quarter of 2011 was $46.0 billion, an increase of 17.5% from the first quarter of 2010 while total retail sales increased 8.6% in the same period. E-commerce sales in the first quarter of 2011 accounted for 4.5% of total sales[1].

The expansion of online commerce has attracted many scholars from diverse disciplines such as Economics, Business and Operation Research, Computer and Information science, Law and others (for a review of e-commerce literature see [16] and [17]).

Trivially, the Web has enabled consumers to access round-the-clock services and to search and compare products, prices, catalogues, descriptions, technical specifications and so forth. Apart from searching and comparing the characteristics of goods and services in the Web, consumers can comment and be informed from others' consumers' purchases and comments. Consumption becomes more connected in the Web. In the rest if this subsection we describe the basic characteristics of connected consumption as a broad economic phenomenon related to recommendation systems.

Basically, positive *network effects* characterize a good when more usage of the good by any User increases its value for other Users. These effects are also called *positive consumption or demand side externalities*. As consumers become more connected in the Web ecosystem, the network effects are gradually based on the mutual benefits of consumption, [18]. *Connected consumption* defines a new form of direct complementarity among consumers. When Users consume goods through the Web, reveal and contribute private information about their preferences and expectations, which is beneficial to other consumers if aggregated and made public. These publicly aggregated consumption patterns and comments are valuable in two ways: (a) indirectly, by reducing search and transaction costs (e.g. tags, playlists, collaborative filtering) and (b) directly, by increasing consumption gains (e.g. discovery of complementary goods in co-purchase networks [19]). Actually, what connected consumers create is not simply content (e.g. reviews) but *context*. This new contextual framework of consumption emerges through the aggregated personal preferences about goods in the Web in massive scale. More importantly, facilitates connected consumers to search and navigate the complex Web more effectively and amplifies incentives for quality. *But how so many and heterogeneous consumers around the globe can coordinate their preferences and expectations?*

---

[1] http://www.census.gov/retail/mrts/www/data/pdf/ec_current.pdf

[2] http://www.cim.co.uk/resources/understandingmarket/definitionmkting.aspx).

In the world of Coase, a small number of consumers can effectively coordinate their preferences through informal agreements and formal contracts to capture the benefits of network effects [20]. However, the coordination of large number of consumers requires high transaction costs. Hayek [21] argued that the price system acts as a coordination device that synchronizes substantial numbers of producers and consumers. Spulber [18] extended Hayek's analysis of "spontaneous order" to include many other market mechanisms for accomplishing coordination at large. These coordination devices include mass media and marketing, mass communications and observation of other consumers.

In the Web era, search engines, social networks and recommendation systems of online retailers (e.g. Amazon, BestBuy) are the most prominent examples of mass coordination devices of consumers' preferences. Search engines and social networks are general coordination devices that include the full spectrum of preferences. Recommendation systems of the Web merchants are focused in increasing the amount of sales by synchronizing purchasing patterns and adapting content provision in individual Users.

It is beyond the scope of the present article to review and analyze the various dimensions of Web economy and commerce, but rather to focus on a specific part of it: the business implications of the technical aspects characterizing recommendation systems.

## 2 Recommendation systems in the Web: the technical aspect

### 2.1 Introduction

Recommendation systems are basic aspects of the current collaborative Web era that complement search engine algorithms in information discovery. Today, almost every Web commerce business uses information filtering techniques to propose products for purchase like a "virtual" salesperson.

Actually, recommendation systems are information filters that exploit user's characteristics (e.g. demographics) and preferences (e.g. views and purchases) to form recommendations or to predict user's future behavior. Commonly, recommendations in the Web are made automatically based on either individual or collective preferences (collaborative filtering) and are presented as hierarchical lists or schemes. For recent surveys on the technical aspects of implementing and analyzing recommendation systems you may refer to [11] and [22].

In the rest of the section we provide a minimal descriptive framework of existing research in recommendation systems that focus on understanding the main characteristics of related functions in the Web. The proposed classification builds on the previous attempts in related research (see for instance Adomavicius and Tuzhilin [23]) and extends the underlying categories by taking into consideration the criteria of input data source and availability. In addition, the concepts of "items" and "products" are specified to the more relevant concept of "Web Goods".

**2.2 Web Goods and Users**

Recommendation systems have emerged to elaborate efficient searching of goods in the Web through personalized and collective evaluation by the Web Users. Before going into the details of recommendation systems let us define what kind of goods are available in the Web.

First, Web Goods has been defined as sequences of binary digits, identified by their assigned URI and hypertext format, and affect the utility of or the payoff to some individual in the economy [24]. Their market value stems from the digital information they are composed from and a specific part of it, the *hyperlinks*, which link resources and facilitate navigation over a network of Web Goods. Web goods can be further elaborated in the following categories. *Pure* Web Goods are the primary focus of the Web Science research [25] because they are defined to include goods that are basically exchanged and consumed in the Web and are not tightly connected to an ordinary good or a service (pre-) existing in the physical world. For instance, a blog entry that comments the market of used cars is a pure Web Good, but a car sales advertisement is not. According to a production incentives-based categorization, Web Goods are discriminated into *commercial* (e.g. sponsored search results) and *non-commercial* (e.g. Wikipedia entries). In contrast to commercial, non-commercial Web Goods are produced outside the traditional market mechanisms of price and property and are based on openness, Peer production and qualitative ex post reward schemes.

In recommendation systems literature, Web Goods are commonly referred as "items" or "products". In the present article we interchangeably use the term of "Web Goods" with the established terms because it better describes the realistic spectrum of goods and services available in and through the Web.

Web Users (or simply Users) produce and consume Web Goods. The emergence of the participatory Web highlighted the decisive role of Web Users in the collaborative creation of online content (refer to Vafopoulos [24] which provides a simple and comprehensive categorization of Web Users based on motivations and economic impact of their actions in the Web ecosystem).

**2.3 The main classifications of recommendation systems**

In the present article we adopt the minimal descriptive definition of recommendation systems initiated by Berkovsky et al [26] in order to enable the comparative analysis of the technological characteristics with the emergent business implications. For more formal and detailed definition of recommendations systems refer to [11].

Initially, it is assumed the existence of $N$ Users with $n$ distinct features, which may request recommendations for $M$ Web Goods with $m$ distinct features. All possible User and Web Good pairs are described by a $n + m$ dimensional space. In the simplest case, a single feature as unique identification describes Users and Web Goods, resulting a two-dimensional space. The $N \: x \: M$ User-Web Good rating matrix represents the ratings given by the Users to Web Goods. These ratings could be formed explicitly or implicitly in a predefined scale. Explicitly is considered in the sense that are directly contributed by Users and not by the recommendation system (implicit).

Table 1: list of symbolic representations of the main variables in recommendations systems' framework

| | |
|---|---|
| $U$ | $1, 2, \ldots, N$ Users with $n$ distinct features |
| $I$ | $1, 2, \ldots, M$ Web goods with $m$ distinct features |
| $User_{feat}$ | the user features (e.g. age, location, income) |
| $User_{id}$ | a unique identifier of the Users |
| $Web \: Good_{feat}$ | Web Good features (e.g. ID, price, availability) |
| $Web \: Good_{id}$ | a unique identifier of the Web Goods |
| $Rating$ | the ratings given by the Users to the Web Goods |
| $R_{gen}$ | the general Recommendation function |
| $R_{CF}$ | the Recommendation function in collaborative filtering |
| $R_{CB}$ | the Recommendation function in content based filtering |
| $Exp_{CA}$ | Context-aware experience |
| $Context_{feat}$ | the context features (e.g. personal attitudes and tasks) |
| **Data Models** | |
| $Rating$ | $Rating \sim (numerical, ordinal)$ |
| $Binary$ | $Rating \sim (0, 1)$ |
| $Unary$ | $Rating \sim (\emptyset, 1)$ |

The general Recommendation function ($R_{gen}$) is described as follows:

$R_{gen}$: User$_{feat}$ x Web Good$_{feat}$ → rating   (1)

Since in most cases (1) is not defined for all possible $n + m$ Users-Web Goods pairs (sparsity problem), a completion rule is needed to fill the missing values. Therefore, the main focus of analysis of recommendation systems is twofold: (a) to estimate the ratings of Web Goods that have not been rated by the Users and (b) to provide methodologies and techniques that will facilitate the formation of recommendations (Table 1 contains the symbols used in this section).

Missing values of the not-yet-rated Web Goods can be assessed by either empirical validation of specific heuristic forms of the recommendation function (1) (data-driven approach) or by estimating the recommendation function that optimizes statistical performance criteria (e.g. MSE) (model-driven approach). For a comprehensive review according to the rating estimation approach refer to [23].

In the present article, recommendation systems are categorized according to two sets of different criteria: how recommendations are created and what kinds of data are available, and a unified analytical framework is provided under common symbolism.

### 2.3.1 How recommendations are formed

Regarding to *how recommendations are formed,* recommendation systems can be classified into the following three categories [27]:

*a. Content-based recommendations*

The User will be recommended Web Goods similar to her past preferences [28]. In the case of content-based recommendations the two-dimensional matrix $R_{CB}$ is given by the following representation:

$R_{CB}$: User$_{id}$ x Web Good$_{feat}$ → rating   (2)

where User$_{id}$ is a unique identifier of the Users, Web Good$_{feat}$ refers to a feature space that represents the Web Good's features and rating reflects the User's evaluation for the Web Good's features [26].

*b. Collaborative recommendations*

The User will be recommended Web Goods that groups of Users with similar tastes preferred in the past (e.g. co-purchase network of Web Goods). In the case of collaborative filtering the two-dimensional matrix $R_{CF}$ is becoming:

$R_{CF}$: $User_{id}$ x Web $Good_{id}$ → rating   (3)

Cacheda et al [29] in their recent work compare different techniques of collaborative filtering by identifying their main advantages and limitations.

*c. Hybrid recommendations*

In the hybrid approach, content-based and collaborative methods are orchestrated in forming recommendations (for a survey see [30]).

Adomavicius and Tuzhilin [23] contributed a more detailed classification of recommendation systems research by analyzing the statistical methodology followed in each of the above main three categories. In particular, they discriminated user-based (or memory - or heuristic - or neighborhood-based) and model-based recommendation techniques for each one of the content-based, collaborative and hybrid approaches. User-based filters consider that each User participates in a larger group of similarly behaving individuals and therefore, physical or Web Goods frequently viewed, liked or purchased, by group members of the group, are the main input for recommendation algorithms [31]. User-based algorithms are heuristics that classify items based on the entire collection of previously rated Web Goods by the Users. The model-based approach *"analyzes historical information to identify relations between different items such that the purchase of an item (or a set of items) often leads to the purchase of another item (or a set of items), and then use these relations to determine the recommended items."* [31]. The most popular type of model-based recommendations in Web commerce is referred to the literature as the "item-based top-N recommendation algorithms" (example is presented in subsection 3.4.4). These algorithms exploit the similarities among various Web Goods to define the set of them to be recommended (for an updated survey on collaborative filters refer to [32] and [11]).

**2.3.2 Data sources and data availability**

The second main category of recommendation systems is based on *data sources* and *data availability*. Regarding the data source, four different types of User's feedback are identified: no feedback, explicit, implicit and hybrid feedback. Explicit feedback is formed by direct input of Users regarding their preferences for specific items. For instance, Amazon reviews; Netflix star ratings and similar high quality data can be used in collaborative filtering algorithms. In cases where explicit feedback is not available or inadequate for building efficient collaborative systems, implicit feedback is employed. Implicit feedback data basically include browsing, usage patterns, purchase history and social network analysis. For instance, usage patterns in hypermedia systems are employed to enable adaptation to the individual User's needs (for a review see [33]). The aforementioned item-based top-N recommendation algorithms are based on implicit feedback mechanisms.

According to Yifan Hu et al [34], collection of implicit data is characterized by the following four main characteristics:

1. The option for Users to express no negative feedback is usually absent.
2. Data is inherently noisy.
3. The numerical values of implicit feedback declare *confidence* and not *preference* as in the case of explicit feedback.
4. Evaluation of recommendation systems based on implicit feedback require updated statistical measures that account for new features such as Web Good availability, competition and dynamic feedback.

Hybrid feedback recommendation systems are jointly exploit explicit and implicit feedback from Users (see for instance [35]).

Analysis of recommendation systems can be also indexed on the basis of *data availability*. According to Bodapati [36] three different types of models are analyzed by the relevant literature, namely: the *ratings*, the *binary* and the *unary* data models. Specifically, the above classification refers to the availability of data in the $N \times M$ User-Web Good rating matrix. In the first case of ratings data model, each User provides explicit feedback by reporting a vote for a subset of Web Goods on a numerical (e.g., 1-6) or ordinal (e.g., like, indifferent or dislike) scale (Table 1). The binary data is considered to be a truncated version of the first model since Users express either a positive or a negative feedback. Purchasing a Web Good or awarding it a rating that meets some threshold could identify positive feedback, commonly recorded as 1. On the contrary, the User declares negative feedback (recorded as 0) if she expresses the intention not to purchase the Web Good or if her rating falls below the threshold. Finally, the *unary* data model is a restricted version of the binary data model because only positive valences are observed. Statistical analysis of item-based top-N recommendations is commonly based on the unary data model.

**2.3.3 Context-aware recommendation systems**

In the introductory part of this article we argue that, through recommendation and feedback systems, what connected consumers create is not simply content (e.g. reviews) but *context*. This new contextual framework of consumption emerges through the aggregated personal preferences about goods in the Web and enables connected consumers to search and navigate more effectively and amplifies incentives for quality in the production of online content.

The investigation in different aspects of contextual information is gaining attention in fields not only technical such as the Semantic Web, data mining, information retrieval and computing, but also in economics and business studies. There are many diverse definitions of context. Context in recommendation systems analysis could be defined to consist of five concrete aspects: environment, personal attitudes, tasks, social and spatiotemporal information [37].

Recently, [38] Adomavicius and Tuzhilin contributed a thorough review of context-aware recommendation systems. They highlight that context-aware recommendations are characterized by complexity and interactivity and they initiate three different algorithmic paradigms for incorporating contextual information into the recommendation process.

In order to capture the various dimensions of feedback and context McNee et al [39] generalized all possible forms of rating to evaluation and Berkovsky et al [26] extended the general Recommendation function $R_{gen}$ to include the experience (Exp) of User for a Web Good. Initially, *"an experience is defined as an evaluation function that maps a pair, the user that had the experience and the item experienced by the user, to an evaluation."* [26]. On this basis, the additional third dimension of context shapes the context-aware experience of the User in a recommendation system. Formally, is represented as follows:

$$Exp_{CA}: User_{feat} \times Web\ Good_{feat} \times Context_{feat} \rightarrow evaluation \quad (4)$$

The inclusion of contextual information into the recommendation process creates new opportunities in personalizing and adapting more efficiently online content through existing and innovative business practices. In the next section, we discuss the marketing implications of recommendation systems based on the understanding of core functional aspects that this section has built.

## 3 Recommendation systems in the Web: the marketing aspect

### 3.1 Introduction

In the current Web, almost every e-commerce business uses information filtering techniques to propose products for purchase like a "virtual" salesperson. Salespersons in the physical world are responsible of making product recommendations to customers, which are integrated and aligned with the firm's marketing strategy. According to the Chartered Institute of Marketing, *Marketing* is the management process, which fulfills the following objectives: identifying, anticipating and satisfying customer requirements profitably[2].

Marketing in the Web (or internet marketing or e-marketing) is intuitively defined as the process of achieving the aforementioned objectives of traditional Marketing through mainly the Web ecosystem [40]. Therefore, recommendation systems in Web 2.0, beyond their role as online "virtual" salespersons, extend firm's marketing strategy by providing a hypermedia two-way channel between producers and distributors and customers. In particular, recommendation systems contribute in the fulfillment of marketing objectives by:

- *Identifying customer requirements*
    - facilitating massive, more detailed and cheaper data acquisition
    - extending one-to-one marketing analysis
- *Anticipating customer requirements*
    - enriching statistical modeling of customer's behavior
    - extending Market Basket Analysis
- *Satisfying customer requirements*
    - providing more informed, personalized and adaptive recommendations
    - implementing one-to-one marketing analysis
    - facilitating better merchandising and atmospherics

---
[2] http://www.cim.co.uk/resources/understandingmarket/definitionmkting.aspx).

**3.2 One-to-one marketing, personalization and adaptation in the Web**

One-to-one marketing (also referred as personalized marketing) instead of targeting an entire group of customers, as in traditional marketing, is designed to increase the revenue of a business by servicing each customer individually and fitting its needs perfectly [41], [42]. Thus, the function of one-to-one marketing is twofold:

- understand each customer's needs individually and
- recommend products that suit the customer's needs

In particular, one-to-one marketing is defined by four principles [43], namely: (1) identify customers, (2) differentiate each customer, (3) interact with each customer and (4) customize products for each customer.

Despite the fact that one-to-one marketing has been employed by researchers and practitioners before the Web, the advances on digital and Web technologies accelerated its expansion. First to mention the contribution of new technology on one-to-one marketing were Gillenson and Sherrell [44]. They also underlined that although super markets dominate the shopping behavior, customers still want to be treated as individuals. They highlighted that *"One-to-one marketing activities are characterized by a desire to interact individually with the most profitable customers of a firm. By learning the needs and desires of the most profitable customers and responding to those desires, companies can build an intensely loyal and profitable clientele."* In the Web era, the massive employment of recommendation systems enables the realization of one-to-one marketing not only to "the most profitable customers" as Gillenson and Sherrell [44] suggests, but for all Web customers. Mainly, one-to-one marketing employs content-based and hybrid recommendations.

One-to-one marketing in the Web converges to the processes of personalization and adaptation, which have extensively investigated in Computer science and related fields. Web personalization could be considered as the process in which the content, the layout and the functionality of a website are being changed dynamically, according to User's features in order to provide recommendations that would fit her. For an extensive discussion concerning various definitions and approaches on personalization refer to [45], [46], [47] and [42].

Kim [42] distinguishes the modes of personalization in two different aspects: information in the Web and one-to-one marketing. However, we suggest that participatory Web commerce through recommendation and feedback mechanisms unifies these aspects. In practice, Web 2.0 merchants attempt to jointly deliver successful marketing mixtures with personalized and adaptive information. Going a step further, online content adaptation could be analyzed as system-driven personalization (not to be confused to adaptability which is User-driven personalization).

### 3.3 Web Merchandising and Atmospherics

Merchandising and store atmospherics are the basic aspects of satisfying the customer requirements, which are included in the third axe of marketing objectives.

In traditional marketing of physical stores, these fields are responsible for efficient placing of products on the "shelf" and creating the appropriate store atmosphere to attract and sustain new customers. The emergence of click-and-mortar commerce resulted important transfers and transformations in the business functions of merchandising and store atmospherics. We briefly discuss the relevant changes to recommendation systems in the Web.

Merchandising *"consists of the activities involved in acquiring particular products and making them available at the places, times, prices and in the quantity to enable a retailer to reach its goals"* [48]. In the Web, there is no physical shelf, retailer or merchandiser but instead websites to present and dispose products (in the form of Web Goods, as have been defined in subsection 2.2). These products are stored in stock centers, homes or databases (in the case of pure Web Goods). *Web Merchandising* focuses on how to make available products in the Web. Online merchandisers are responsible for product collection and display, including promotions, cross-selling and up-selling. Studies in Web Merchandising could be divided in four areas [49]: (1) product assortment, (2) merchandising cues, (3) shopping metaphor and (4) Web design features.

Merchandising cues are techniques for presenting and/or grouping products to motivate purchase in online stores. A good example of merchandising cues is recommendation systems apart from traditional promotion methods [50]. Web design features share similar functionalities and analysis with Web atmospherics.

*Web Atmospherics* is the conscious designing of Web environments to create positive effects in Users in order to increase favorable customer responses. Just like retailers provide important information through atmospherics in conventional stores, online retailers also create an atmosphere via their website, which can affect shoppers' image and experience in the online store [51]. For a more extended overview on Web atmospherics refer to [52], [53] and [54]. For example, Nanou et al [55] investigate the effects of recommendations' presentation on customer's persuasion and satisfaction in a movie recommender system and they concluded that the most efficient presentation method is based on the "structured overview" and the "text & video" interfaces.

Recommendation systems could be employed not only for the product's placement, promotion and related functions but to improve the online environment and atmosphere via dynamic and adaptive features to the User's and product's characteristics. Mass merchants in the Web (e.g. Amazon) use various forms of recommendations (both content-based and collaborative) to build the main part of their store's architecture, functionality and adaptation. Indicatively, the webpage of an item in Amazon[3] contains recommendations of the following types: "Frequently Bought Together", "What Other Items Do Customers Buy After Viewing This Item?", "Customers Who Bought This Item Also Bought", "Customers Viewing This Page May Be Interested in These Sponsored Links", "Product Ads from External Websites", "Customer Reviews", "Related Items", "Tags Customers Associate with This Product", "Customer Discussions", "Listmania!", "Recently Viewed Items" and "Recent Searches".

To conclude, the transition from traditional marketing to marketing in the Web is not a trivial task. Recommendation systems contribute in this transition as follows:

- Marketing objectives sustain their core principles but extend their field of implementation in a more interactive communication process between business and customers and among customers themselves.
- In this updated communication process, recommendation systems play a prominent role by aggregating individual preferences and enabling massive one-to-one marketing.
- Recommendation systems also facilitate personalization and adaptation of Web commerce, which drastically affect Web merchandising and atmospherics.

We suggest that one-to-one marketing, recommendation systems and adaptation in the Web are crucial functions that should be further explored, both in Marketing and Web studies, in order to enhance Web merchandising and atmospherics.

---

[3] The considered example refers to Apple MacBook Pro MC725LL/A 17-Inch Laptop http://www.amazon.com/Apple-MacBook-MC725LL-17-Inch-Laptop/dp/B002C74D7A/ref=sr_1_5?s=pc&ie=UTF8&qid=1309243292&sr=1-5

### 3.4 Market basket analysis

### 3.4.1 Introduction

There is a classic example of Market Basket analysis stating, "beers and diapers are often being purchased together in the same basket". In the Web, beers and diapers have been expanded to online music, movies and various types of services. The new marketing strategy mix has to anticipate changes in purchase behavior and customer requirements. Market Basket analysis is a prominent tool in this effort since a growing number of research communities are involving in understanding consumption patterns in Web commerce.

Apart from traditional database and data mining-oriented research on Market Basket analysis, investigators in network analysis and econometrics have recently contributed rich insights in this field. Network analysis is mainly based on crawling publicly available data from recommendation systems in the Web (e.g. Amazon).

### 3.4.2 Definitions, history and applications

Let us first describe the basic definitions and techniques of Market Basket analysis.

*Market Basket* is the set of items purchased by a customer during one single shopping occasion [56]. During a shopping trip, the customers are in a "pick-any"-situation because they have the option to choose no item, one or any other number of items from each category [57]. Market Basket data are binary data (e.g. an item added or not added into basket) organized in sets of items bought together by customers (often called transactions [58]).

*Market Basket Analysis (MBA)* studies the composition of shopping baskets in order to identify customer's purchase behavior [56]. It is also known as *association rule mining*, which is a method of identifying customer purchasing patterns by extracting associations from stores' transactional databases [59].

A mathematical definition of MBA is provided by Chen et al [59]: *"Given two non - overlapping subsets of product items, X and Y, an association rule in form of X|Y indicates a purchase pattern that if a customer purchases X then he or she also purchases Y. Two measures, support and confidence, are commonly used to select the association rules. Support is a measure of how often the transactional records in the database contain both X and Y, and confidence is a measure of the accuracy of the rule, defined as the ratio of the number of transactional records with both X and Y to the number of transactional records with X only."*

MBA is occupied for mainly marketing purposes. In particular, could be helpful in designing and implementing:

- cross and up-selling strategies[4]
- promotion strategies and discounts
- loyalty programs
- store atmospherics

Chib et al [60] summarize the motivation of MBA in the following arguments:

- Create improved estimates of brand-choice elasticities with respect to marketing mix variables, properly accounting for not just the direct impact but also the indirect impact on brand-choice via category purchases [61]
- Facilitate the understanding of what factors drive category purchase incidence and what impact marketing-mix variables have at the brand level on category purchases.
- Describe the isolating correlations amongst various product categories within the shopping basket in order to identify which categories are complements and which are substitutes.

Before the advent of the Web, MBA was enhanced by the technological evolutions on transactional systems in commerce (e.g. barcode implementations, RFID etc.) (see for example [62] and [56]). Electronic transactions provided the first stream of massive Market Basket data. Aggarwal et al [62] were the first to propose an influential algorithmic way for data mining in large transactions databases. One of the tools commonly used to perform MBA is Affinity analysis. Affinity Analysis is a technique that discovers co-occurrence relationships among transactions performed by specific individuals or groups. For a literature review on MBA and related techniques refer to Mild and Reutterer [56] who classify relevant literature depending on the followed statistical approach (exploratory or explanatory analysis).

The second stream of input data came from Web commerce transactions including clickstreams, log files and other browsing data captured through voluntary and compulsory collection processes.

Hao et al [63] argue that MBA has become a key success factor in e-commerce and Kantardzic [64] states *"A business can use knowledge of these patterns to improve the Placement of these items in the store or the layout of mail-order catalog page and Web pages."*

---

[4] Cross-selling and up-selling are strategies of providing existing customers the opportunity to purchase additional or more expensive items, respectively.

### 3.4.3 Market Basket Analysis and recommendation systems

MBA and recommendation systems in Web commerce share the same input transaction data but employ different techniques and address their results to different end users. MBA identifies customer's purchasing patterns by extracting associations from the data to inform the decisions of marketing managers, while recommendation systems provide relative information to Web Users.

Traditional MBA in purchasing data of physical stores is an exclusive property of storeowners and contains less information about the customer's purchasing behavior than in online stores. Specifically, in Web commerce every purchase is assigned with a unique time stamp of occurrence, Users' reviews and evaluations are often contributed and purchasing behavior could be inter-connected to general browsing patterns and website visits. These extra features of commerce data analysis in the Web offer a comparative advantage to online merchants and raises concerns of excessive market power and personal data abuse through selling to third parties and profiling without permission.

What is also changing in Web commerce is that other entities than the store owner/administrator can commence (partial) MBA by collecting online recommendations.

For example, Amazon, the biggest Web merchant, is based on a successful item-based collaborative filtering system[5] providing a wide range of general and personalized recommendations. Specifically, the list of "most customers that bought this item, also bought" (BLB) recommended products presents related items that were co-purchased most frequently with the product under consideration.

BLB recommendations form the store's co-purchase network and can be represented as a directed graph in which nodes are products and directed links connect each product with its recommended products. In such setup: *"The virtual aisle location of a product is determined, in part collectively by consumers rather than being chosen based on fees paid by manufacturers, or explicit strategic considerations by the retailer"* [65].

### 3.4.4 Network analysis of Market Basket data

The idea of examining the Amazon BLB co-purchase network was initiated by Krebs [66] who proposed the analysis of emergent patterns of connections that surround an individual, or a community of interest, based on book purchases. Dhar et al [67] extended considerably the analysis by assessing the influence of BLB networks on demand and revenue streams in Web commerce. They also contributed new experimental verification about the significance of visible item recommendations on the long tail of commerce in the Web.

---

[5] For more detailed descriptions of collaborative filtering please refer to subsection 2.3.1.

Oestreicher-Singer and Sundararajan conducted a series of research efforts to analyze time series of co-purchase item networks incorporating methods from economic theory, econometrics and computer science. By repeatedly crawling the same items they create time series data for the books under consideration. In their initial research Oestreicher-Singer and Sundararajan [68], [69], they employed a depth-first crawler to collect from Amazon approximately 250,000 distinct books during the period 2005-6.

Their main focus was to infer demand levels for each item and to evaluate if the network structure influences individual items. This influence was measured by adapting the PageRank algorithm to account for weighted composite graphs. The variation in the demand distribution across categories was estimated by computing the Gini coefficient [70] for each category. Their conjecture was found to be in accordance to the "long tail" for demand phenomenon in ecommerce [71]. Their results also include the following: *"(1) an increase in the variance in the extent to which the network influences products in a within a category increases the category's demand inequity. (which makes intuitive sense in the context of our theory of the network "flattening" demand), (2) The number of products in a category is positively associated with demand inequity, and (3) the average demand within a category is associated with an increase in the category's demand inequity."*

Using a similar to [68],[69], dataset Oestreicher-Singer and Sundararajan [72] econometrically identified the influence that visible co-purchase networks have on demand of individual items. Based on a simple set of conditions, which imply minimal empirical restrictions on the network structure concluded that visible co-purchase networks more than triples the average influence that complementary items have on demand. They also estimated that the magnitude of this social influence is higher for more popular and more recently published books. On the contrary, pricing, secondary market activity and assortative mixing across product categories are related by counter-intuitive ways with network position of individual items. Furthermore, in a newer investigation [65] they found among others that within a books category an increase in the influence of the recommendation network is consistently related with a more even distribution of both revenue and demand and when the recommendations are internal to a category itself, *"the redistribution of attention they cause compensates demand more within the category, rather than redirecting demand to a popular book in a different category."*

These results have practical implications for Web commerce since it is becoming clear that connected consumption exists online.

In 2009, Carmi et al [73] extend [72] to address new research questions related to the diffusion of exogenous shocks in the Amazon co-purchase network. In particular, they estimated how far these shocks propagate, how long they survive and if they affect demand of neighboring items and network structure. The initial dataset was augmented to include two years more data and book reviews about the books of Oprah Winfrey that appeared on the Oprah.com and "Sunday Book Review" section of the New York Times. They identified a consistent relationship between the shape of the diffusion curve and the level of clustering in the co-purchase network.

They also concluded that two subsets of reviewed books exist: "those whose demand increase is substantially higher than the total increase for its neighbors, and those for which the total increase in demand for the neighbors is an order of magnitude higher."

Going a step further from the econometric identification of impact that the visible co-purchase links have to demand [65], Dhar et al [67] assessed whether co-purchase item networks contain useful predictive information about sales. In particular, they estimated a simple auto-regressive model, where demand in the next period is modeled as a linear combination of demands in previous periods. By analyzing an extensive dataset, which covers a diverse set of books spanning over 400 categories over a period of three years with a total of over 70 million observations they concluded that changes in demand for each item can be predicted more accurately using network information.

Recently, in a relevant experimental study, Vafopoulos et al [19] crawled a set of 226,238 products from all the thirty Amazon's categories, which form 13,351,147 co-purchase connections. They introduce the analysis of local (i.e. dyads and triads) and community structures for each category and the more realistic case of different product categories (market basket analysis). Their main results concerning the purchasing behavior of Web consumers are the following:

- The cross-category analysis revealed that Amazon has evolved into a book-based multi-store with strong cross-category connections.
- Co-purchase links not only manifest complementary consumption, but also switching among competitive products (e.g. the majority of consumers switch from Kaspersky to Norton Internet security suite).
- Top selling products are important in the co-purchase network, acting as hubs, authorities and brokers (or "mediators") in consumer preference patterns.
- Ostensibly competitive products may be consumed as complements because of the existence of compatibility and compatible products that facilitate their joint consumption.

Let us focus on the community analysis of the co-purchase network of products. For a given network, a community (or cluster, or cohesive subgroup) is defined to be a sub-network whose nodes are tightly connected, i.e. cohesive. Since the structural cohesion of the nodes can be quantified in several different ways, many different formal definitions of community structures have been emerged [74]. The analysis of community structures offers a deeper understanding for the underlying functions of a network. Figure 1 shows a part of the software products co-purchase network, where different colors indicate different community membership. Different product communities have been identified based on the spin glass community detection algorithm [75].

Analysis indicated that the seemingly competitive products of Apple and Microsoft are in reality consumed as if they were complementary. Microsoft (nodes with purple color) and Apple (nodes with orange color) product communities are "mediated" by compatibility like VMware Fusion, Parallels Desktop and compatible products like Office for Mac.

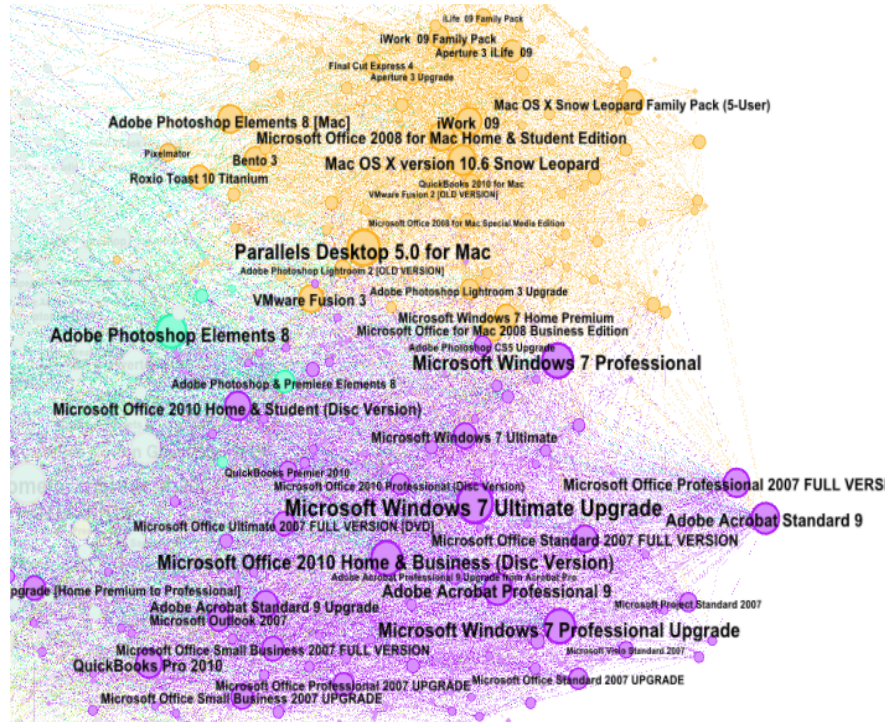

Fig. 1 Microsoft and Apple software programs are consumed as complements, because of compatibility (e.g. Parallel Desktops) and compatible (e.g. MS Office for Mac) products.

**3.5 Network-based marketing**

In parallel, apart of e-marketing studies has been recently emerged the field of network-based marketing. Network-based marketing refers to a collection of marketing techniques that take advantage of links between consumers to increase sales and should not be confused with network or multilevel marketing [76]. It could be also found in the literature as word-of-mouth and buzz marketing [77] and viral marketing [78]. The main focus is to measure how product adoption propagates from consumer to consumer through recommendation systems [78] and customer feedback mechanisms [79], [80].

Contrastingly to traditional marketing studies, network-based marketing models interdependent consumer preferences through explicit and implicit links among consumers. According to Hill et al [76] statistical research in network-based marketing includes six main fields: (1) econometric modeling, (2) network classification modeling, (3) surveys, (4) designed experiments with convenience samples, (5) diffusion theory and (6) collaborative filtering and recommendation systems.

Recommendation systems are relevant to network-based marketing because share the same objective to exploit the underlying knowledge residing in the stored data that are related to customer behavior (for a review in related literature refer to [81]). As Hill et al [76] highlight *"Recommendation systems may well benefit from information about explicit consumer interaction as an additional, perhaps quite important, aspect of similarity."*

Leskovec et al [78] studied an extensive snapshot of Amazon's person-to-person recommendation network of products. They modeled propagation of recommendations and the cascade sizes according to a simple stochastic model and concluded that product purchases follow a 'long tail' and that on average recommendations are not very effective at inducing purchases. Based on Bayesian network analysis identified communities, product, and pricing categories for which viral marketing is considered to be efficient.

## 4 Discussion

Making good product recommendations in the Web is not just matter of fast algorithms, but also a business task. Moreover, selling products and services in the Web has become a complex issue with equally important technical and business aspects, because Users, apart from searching and comparing the characteristics of products, can comment and be informed from others' Users' purchases and comments. In Web 2.0, Users reveal and contribute private information about their preferences and expectations, which is beneficial to other consumers if aggregated and made public. Consumption becomes more connected in the Web. This fact calls for new ways of investigating related Web phenomena and behavior.

In this article, we review recommendation systems, not only as technical artifacts, but also as parts of the more general problem of studying online purchasing behavior. The main focus is to highlight useful connections among diverse research efforts, which share some common tasks and challenges. In particular, we discuss the specific fields of one-to-one marketing, network-based marketing, Web merchandising and atmospherics and their implications in the processes of personalization and adaptation in the Web. The transformation of traditional marketing methodologies in the Web ecosystem is a multifold task, in which recommendation systems could contribute because (a) Marketing objectives sustain their core principles but extend their field of implementation in a more interactive communication process between business and customers and among customers themselves, (b) in this updated communication process, recommendation systems play a prominent role by aggregating individual preferences and enabling massive one-to-one marketing and (c) recommendation systems also facilitate personalization and adaptation of Web commerce, which drastically affect Web merchandising and atmospherics.

Market basket analysis is analyzed in the context of recommendation systems. The discussed issues in related literature are micro-economic issues, which directly refer to what types of recommendations are appropriate in achieving certain tasks. But participatory and interactive Web commerce raises a series of issues related to the Web economy in the macro level. As more and more companies are participating in the Web commerce, finding and analyzing consumption patterns is an essential key to their success. Data is the "king" and trust the "queen" in Web commerce and if are combined with navigational patterns and social networking, give to online mass merchants a strong comparative advantage, not only against their direct competitors in the Web but also against to the "brick-and-mortar" retailers.

At the same time, these massive amounts of personal and market data raise concerns about *privacy* and *excessive market power*. To the best of our knowledge, there is no yet scientific investigation in the economic or law literature concerning the excessive market power of Web merchants, which steams from data exploitation. As Clemons and Madhani [82] admit: *"Some digital business models may be so innovative that they overwhelm existing regulatory mechanisms, both legislation and historical jurisprudence, and require extension to or modification of antitrust law."*

An emergent challenge for recommendation systems will be the case of extensive application of Web 3.0 technologies in Web commerce (e.g. Good Relations ontology [83]). The easier exploration and comparison of Web Goods from Users will enable further competition and lower price dispersion. Semi-automatic contracting and business rules formation [84] have the potential to extend recommendation systems in a wider range of functionality.

The research agenda of economists, computer and information scientists is filling up with issues coming from Web commerce and in our point of view, this is going to happen for many years. Experts in algorithms, statistics and business intelligence will experiment with alternative recommendation and feedback systems (e.g. context-aware systems) but it seems that their efforts should account more seriously for data privacy concerns. An alternative and more generic approach may be to extend the Web architecture to support ex ante information transparency and accountability rather than ex post security and access restrictions [85]:

*"Consumers should not have to agree in advance to complex policies with unpredictable outcomes. Instead, they should be confident that there will be redress if they are harmed by improper use of the information they provide, and otherwise they should not have to think about this at all".*

Consequently, future challenges in recommendation systems will not be purely technical or business-oriented. They will involve issues like privacy, trust and provenance in semantic and ubiquitous Web environments, market competition and regulation.